\documentclass[prb,twocolumn,amsmath,amssymb,floatfix,superscriptaddress]{revtex4}
\usepackage{graphicx}
 
\newcommand{\be}{\begin{eqnarray}}
\newcommand{\ee}{\end{eqnarray}}
\newcommand{\bfr}{{\bf r}}

\newcommand{\bfp}{{\bf p}}

\newcommand{\bfE}{{\bf E}}
\newcommand{\bfd}{{\bf d}}
\newcommand{\bfe}{{\bf e}}
\newcommand{\bfJ}{{\bf J}}

\newcommand{\wbe}{\begin{widetext}}
\newcommand{\wee}{\end{widetext}}
\newcommand{\oncite}{\onlinecite}

\begin{document}

\title{Field Induced Long-lived Super-Molecules}


\author{S.-J. Huang}
\altaffiliation{
These authors contributed equally to the work.
}
\affiliation{Physics Department and Frontier Research Center on Fundamental and Applied Sciences of Matters, National Tsing-Hua University, Hsinchu, Taiwan}
\affiliation{Physics Division, National Center for Theoretical Sciences,
Hsinchu, Taiwan}

\author{Y.-T. Hsu}
\altaffiliation{
These authors contributed equally to the work.
}
\affiliation{Physics Department and Frontier Research Center on Fundamental and Applied Sciences of Matters, National Tsing-Hua University, Hsinchu, Taiwan}
\affiliation{Physics Division, National Center for Theoretical Sciences,
Hsinchu, Taiwan}

\author{H. Lee}
\affiliation{Physics Department and Frontier Research Center on Fundamental and Applied Sciences of Matters, National Tsing-Hua University, Hsinchu, Taiwan}
\affiliation{Physics Division, National Center for Theoretical Sciences,
Hsinchu, Taiwan}

\author{Y.-C. Chen}
\affiliation{PInstitute of Atomic and Molecular Sciences, Academia Sinica, Taipei, Taiwan}

\author{A. G. Volosniev}
\affiliation{Department of Physics and Astronomy, Aarhus University, Aarhus C, DK-8000, Denmark}

\author{N. T. Zinner}
\affiliation{Department of Physics and Astronomy, Aarhus University, Aarhus C, DK-8000, Denmark}

\author{D.-W. Wang}
\affiliation{Physics Department and Frontier Research Center on Fundamental and Applied Sciences of Matters, National Tsing-Hua University, Hsinchu, Taiwan}
\affiliation{Physics Division, National Center for Theoretical Sciences,
Hsinchu, Taiwan}


\date{\today}

\begin{abstract}
We demonstrate that the long-lived bound states (super-molecules) can exist in the dilute limit when we tune the shape of effective potential between polar molecules by an external microwave field. Binding energies, average sizes, and phase diagrams for both $s$-orbital (bosons) and $p$-orbital (fermions) dimers are studied, together with bosonic trimer states. We explicitly show that the non-adiabatic transition rate can be easily tuned small for such ground state super-molecules, so that the system can be stable from collapse even near the associated potential resonance. Our results, therefore, suggest a feasible cold molecule system to investigate both novel few-body and many-body physics (for example, the $p$-wave BCS-BEC crossover for fermions and the paired condensate for bosons) that can not be easily accessed in single species atomic gases.
\end{abstract}

\maketitle
Since the successful realization of a high-phase-space density of molecules at JILA [\oncite{JILA}], systems of ultracold polar molecules become one of the most exciting developments in strongly correlated physics. However, the strong dipolar interaction between polar molecules may render the system unstable to collapse in the high density regime (also observed in atomic gases with negative scattering length). Recently there have been several methods proposed to overcome such an unwanted situation by reducing the system dimensions in an optical lattice [\oncite{wang,others_layer,shlyapnikov,interlayer_superfluid,quasi_1D,bound_state}]. 
On the other hand, an effective repulsive core can also be generated to stabilize the system if a proper detuned AC field is applied [\oncite{zoller}]. These mechanism open new possibilities to investigate interesting many-body physics of polar molecules, including the $p$-wave superfluid between fermionic molecules [\oncite{shlyapnikov}], the itinerant ferro-electricity [\oncite{ferro_electricity}], and dimerized superfluidity [\oncite{Andrew}] etc.

\begin{figure}
\includegraphics[width=8cm]{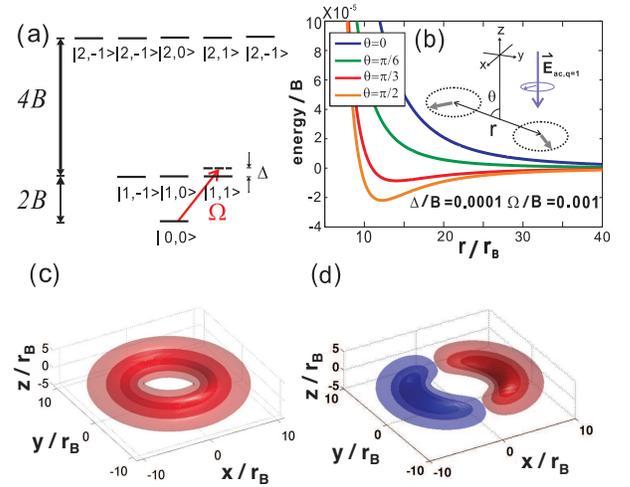}
\caption{(Color online) (a) Rotational eigenstates of a single molecule. The microwave field couples states $|0,0\rangle$ and $|1,1\rangle$ with the Rabi frequency $\Omega$ and the detuning $\Delta$ (circularly polarized in the $x-y$ plane as shown in the inset of (b)). (b) The effective interaction ($V_{\rm eff}(r)$) for different inclination angles $\theta$. $\theta=0$, $\pi/6$, $\pi/3$, and $\pi/2$ respectively from top to bottom. (c) and (d) show the 3D contour plots of the ground state wavefunction of a dimer in the $s$- and the $p_x$-state respectively. Red(blue) color stands for positive(negative) value of the wavefunction. Here we use $\Omega/B$=0.01 and $\Delta/B$=0.01 for SrO molecules.}
\label{system}
\end{figure}

In addition to many-body physics, it is also interesting to study the few-body physics of polar molecules in the dilute limit. In Ref. [\oncite{bohn}], the authors proposed a DC-field linked bound state by pumping the electrons to a meta-stable state, which, however, has a very short life time and is therefore not easily accessible. DC field induced inter-layer(-tube) bound states in a multi-layer(-tube) system were also proposed [\oncite{bound_state}], but the mutual long-ranged repulsion makes the critical density for such bound state formation much lower than the density available in current experiments [\oncite{bilayer_Zinner}]. 

In this paper, we investigate a different few-body bound state (super-molecule) in the presence of an external AC field [\oncite{Gora_work}]. The effective potential between polar molecules can be tuned to have an attractive and shallow well (see Fig. \ref{system}(a) and (b)), which supports only a few or even no bound state (dimers and trimers) between molecules. Their average sizes are of sub-micrometer scale, strongly suppressing the few-body inelastic scattering rate and stabilizing the whole system from collapse. As a result, our work suggests a new area for novel few-body physics with a long-lived ground state. Some important many-body physics not accessible in atomic gases, for example the $p$-wave BCS-BEC crossover between single species fermions [\oncite{fermion_pwave}] and the $Z_2$ transition [\oncite{Z2}] between the condensate and the pair condensate of single species bosons near the Feshbach resonance, may become feasible without particle loss. 


In the low temperature and dilute limit, the only important internal degrees of freedom is the rotational states, if considering close-shell molecules (SrO as an example) without hyperfine structure. As a result, the general two-body Hamiltonian with dipolar interaction becomes [\oncite{zoller}]: 
$H=\sum_{i=1,2} \left[\frac{\bfp_i^2}{2m}+B\bfJ_i^{2}- \bfd_i\cdot \bfE(t)\right]+\frac{\bfd_1\cdot \bfd_2
-3(\bfd_1\cdot\bfe_{12})(\bfd_{2}\cdot\bfe_{12})}
{|\bfr_1-\bfr_2|^{3}}$,
where $m$, ${\bf J}_i$ and $B$ are the molecular mass, angular momentum operator and the rotational constant respectively. $\bfE(t)$ is the external field, and $\bfd_i$ is the electric dipole moment operator. $\bfr_i$ and $\bfp_i$ are the position and momentum operators. $\bfe_{12}\equiv \bfr_{12}/|\bfr_{12}|$ and $\bfr_{12}\equiv\bfr_1-\bfr_2$. Here we will consider only the case when a circularly-polarized microwave field is tuned to be close to the transition energy between $|0,0\rangle$ and $|1,1\rangle$ (see Fig. \ref{system}(a)). Here $|J,M\rangle$ is the rotational state labeled by the angular momentum quantum number. 

Using the adiabatic and rotating wave approximations [\oncite{zoller}],
we obtain the full effective interaction [\oncite{shlyapnikov}], which has the leading order terms in the long distance:
\be
V_{\rm eff}(\bf{r})&=& -\frac{d_{\rm eff}^2}{|\bfr|^3}(1-3\cos^2\theta) \\
&&+\frac{\Delta}{2\Omega^2}\frac{d_{\rm eff}^4}{|{\bfr}|^6}\left[C_6^{(1)}(\theta)+C_6^{(2)}(\theta)\right]
+{\cal O}(|\bfr|^{-9}),
\nonumber
\label{V_large_r}
\ee
where $d_{\rm eff}\equiv\frac{d\Omega/|\Delta|}{\sqrt{3(1+4(\Omega/\Delta)^2)}}$ with $d$ being the bare electric dipole moment. $C_6^{(1)}(\theta)\equiv\frac{(1+3\xi)(1+(\Omega/\Delta)^2)}{3+\xi+12(\Omega /\Delta )^2}(1-3\cos^2\theta)^2$ and $C_6^{(2)}(\theta)\equiv\frac{36\xi^3}{3+\xi+12(\Omega /\Delta )^2}\sin^4\theta$ are two coefficients with $\xi\equiv\sqrt{1+4(\Omega/\Delta)^2}$. $\theta$ is the angle between the microwave propagation and the relative position of the two molecules (see the inset of Fig. \ref{system}(b)). We let $\hbar\equiv 1$.
We can see that $V_{\rm eff}(\bf{r})$ is dominated by the dipolar-like term with the effective dipole moment, $d_{\rm eff}$, as $|\bfr|\to\infty$, but its next order correction is like a hard-core potential to prevent possible chemical reactions [\oncite{zoller,shlyapnikov}]. Therefore, one can always expect a potential minimum at $|\bfr|=r_m$ in the $x-y$ plane.

\begin{figure}
\includegraphics[width=8.5cm]{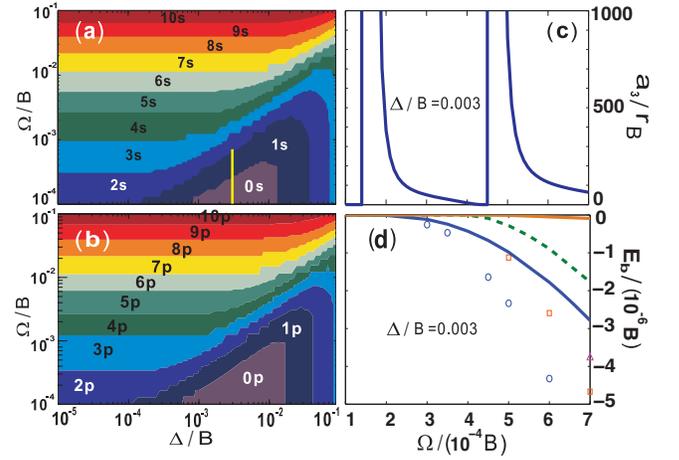}
\caption{(Color online) (a) and (b) are phase diagrams for the number of $s$- and $p$-orbital states of a dimer super-molecule. (c) is the resonance of the $s$-wave scattering length ($a_3$, see text) as a function of $\Omega$ (i.e. the yellow vertical line in (a)). (d) is the corresponding binding energies for the $s$-orbital ground (blue solid line) and the first excited (orange solid line) states. Results for the $p$-orbital ground state, the bosonic trimer ground, the first excited and the second excited state are shown in green dashed line, blue circles, orange squares and purple triangles respectively.}
\label{phase_diagram}
\end{figure}
In Fig. \ref{system}(b), we show the full numerical $V_{\rm eff}(\bfr)$ for different $\theta$ with the detuning $\Delta/B=10^{-4}$ and the Rabi frequency $\Omega/B=10^{-3}$. Note that, by using $r_B\equiv(\frac{d^2}{B})^{1/3}$ and $B$ as the unit of length and energy, $V_{\rm eff}(\bfr)$ is universal for all molecules, once the external field parameters ($\Omega/B$ and $\Delta/B$) are specified. 
Using SrO as an example ($d=8.9$ Debye $B\sim 10^{10} $Hz), we find $r_B\sim 11$nm and hence the super-molecules can be as large as 100 nm as shown in Fig. \ref{system}(c) and (d). The dimer super-molecule has a donut-type wave function for bosons ($s$-orbital state) and has two knots for fermions in the $p_x$-orbital state [\oncite{Gora_work}].

Since all the important and non-trivial properties of the effective potential lies on the $x-y$ plane, from now on, we will consider 2D molecular motion for simplicity and use $V_{\rm 2D}(r)\equiv V_{\rm eff}(|\bfr|,\theta=\pi/2)$ as the effective interaction. Such system can be physically realized by adding a strong confinement in the $z$ direction. The finite width correction can be neglected because the donut type super-molecule has a very small thickness (less than $8 r_B\sim 88$ nm even for SrO).

In Fig. \ref{phase_diagram}(a) and (b), we show the phase diagrams for the number of 2D dimer states in terms of the detuning and Rabi frequencies, including both $s$- and $p$-orbitals. In the limit of small detuning (say $\Delta/B<10^{-4}$), the bound state is solely determined by $\Omega$, while in the limit of large detuning, it depends on $\Delta$ only. In the intermediate regime, there is a triangular window ($0.001<\Delta/B<0.01$), where no dimer bound state exists in both orbital. 

The appearance of a two-body bound state implies a potential resonance in the low energy scattering amplitude. For our 2D effective potential with a $r^{-3}$ tail in large distance, the low energy pseudo-potential ($\hat{\cal V}_{\rm ps}$) can be obtained to be (see Ref. [\oncite{pseudo_wang}] for details):  
\be
\hat{\cal V}_{\rm ps}\psi(\bfr_\perp)&=&\delta(\bfr_\perp)
\frac{2\pi}{2\mu\ln(ka_3/2\beta_0)}\left[\ln\left(\frac{kr}{2\beta_0}\right)\right]^2
\nonumber\\
&&\times
r\frac{\partial}{\partial r}\left[\frac{\psi(r)}{\ln(k r/2\beta_0)}\right],
\label{V_ps_alpha}
\ee
where $\psi(r)$ is the relative wavefunction, $\beta_0\equiv e^{-\gamma}$ and $\gamma\approx 0.57722$ is Euler's constant. Here $a_3\equiv m d_{\rm eff}^2(\beta_0)^{-2}e^{-2/\tilde{P}_0}$ is the effective scattering length [\oncite{pseudo_wang}] with $\tilde{P}_0$ being the only parameter determined from the short-range effective interaction (see Refs. [\oncite{pseudo_wang}] for details). The incident momentum $k$ is related to the system density. The 2D pseudo-potential is valid for $ka_{3}\ll 1$.

In Fig. \ref{phase_diagram}(c), we show the calculated $s$-wave scattering length, $a_3$, as a function of the Rabi frequency at $\Delta/B=0.003$. The corresponding binding energies are also shown in Fig. \ref{phase_diagram}(d). As expected, the divergence of the scattering length (similar to Feshbach resonance) indicates the appearance of an $s$-orbital bound state with a binding energy typically smaller than one $\mu$K (taking $B\sim 10$ GHz). The tunability of the scattering length will certainly affect the many-body properties, and this result remains true for the 3D system. 

\begin{figure}
\includegraphics[width=8.5cm]{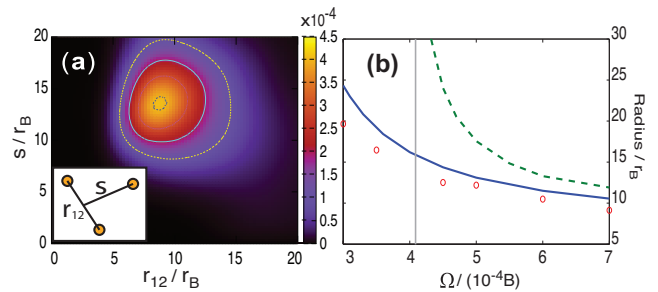}
\caption{(Color online) (a) The wavefunction of a trimer state with $\Delta/B=0.003$ and $\Omega/B=7\times 10^{-4}$ with the coordinates defined in the inset: $r_{12}$ is the distance between two particles, and $s$ is the distance between the third and the center of mass of the first two. Their relative angle has been integrated out. (b) The sizes of bosonic trimer (red circles), $s$-orbital and $p$-orbital dimers (blue solid and green dashed lines respectively) as a function of $\Omega$ for $\Delta/B=0.003$. The vertical line indicates the appearance of the $p$-orbital dimer.
}
\label{phase_trimer}
\end{figure}

In addition to dimer states, we also investigate three-body bound states (trimers). We note that the Efimov related physics in 3D has recently been investigated for dipolar atoms [\oncite{Greene}] or 2D systems [\oncite{Kartavtsev}] with short-ranged interaction. However, the effective interactions of our system are not simple dipolar-like, but with several intrinsic length scales. Therefore one should not expect universal results in our current system. Here we apply the stochastic variational method (SVM) with Gaussian function basis to study trimer states of bosonic molecules [\oncite{varga}]. The uncertainty of our calculation is about three orders of magnitude smaller than the binding energies. In Fig. \ref{phase_trimer}(a) and (b), we show the typical trimer wavefunction and its size as a function of the Rabi frequency.

In Fig. \ref{phase_diagram}(d), we also show the binding energy of a typical trimer super-molecule as a function of Rabi frequency. If we stay close to the threshold of bound state formation, there is a finite parameter range ($2 \times 10^{-4} \lesssim \Omega/B \lesssim 4 \times 10^{-4} $)  in which the energy differences from dimer's are less than $10^{-6}B\sim 10k$ Hz for SrO, smaller than the typical trapping frequency. In other words, even though two super-molecules (four particles) may have inelastic scattering, leading to the formation of one trimer and one free particle, the energy gain of the later should not be high enough to escape the trap. As a result, we do not expect a significant loss rate from such four-body (two super-molecule) scattering, and the resulting ground state should be a mixture of dimer and trimer super-molecules. This is a novel few-body system composed of field-dressed polar molecules. For molecules with smaller dipole moment, $r_{B}$ is smaller. The stable regime of mixture would be enlarged. When we move away from the threshold, the excited states start to show up (at $\Omega/B \sim5\times10^{-4}$ with $\Delta/B=0.003$) with energy lower than a dimer. 

\begin{figure}
\includegraphics[width=8.5cm]{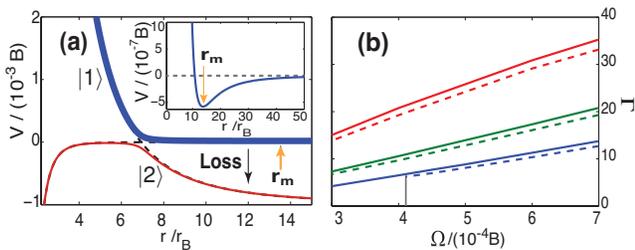}
\caption{(Color online) (a) shows the non-adiabatic transition of a dimer state. The thick blue(thin red) lines are eigenstates $|1\rangle$($|2\rangle$). Dashed lines are states as $\Omega\to 0$. (b) The calculated transition parameter, $\Gamma$, for the $s$-orbital (solid line) and $p$-orbital (dashed line) dimer ground state. $\Delta/B=0.001$, $0.002$, and $0.003$ respectively from top to bottom. The vertical line indicates the appearance of the $p$-orbital dimer for $\Delta/B=0.003$.
}
\label{loss}
\end{figure}

Now we consider the loss rate of a super-molecule due to the breakdown of the adiabatic approximation. The effective potential is made of the avoided crossing between two eigenstates in the weak field limit (see Fig. \ref{loss}(a)) [\oncite{zoller,fermion_pwave}]. They can be denoted to be $|1\rangle$ (effective potential) and $|2\rangle$, merge to $|g,g,N\rangle$ and $|g,e_+,N-1\rangle$ respectively as $r\to\infty$. Here $|g,g,N\rangle$ denotes the two molecules in the rotational ground state with the photon number $N$, while $|g,e_{+};N-1\rangle$ denotes the symmetric superposition of one molecule in the ground state and the other one in the excited state with one photon absorption. The non-adiabatic loss thus mainly from the inter-state transition from $|1\rangle$ to $|2\rangle$ due to the relative kinetic energy between two particles neglected before (see Fig. \ref{loss}(a)). The transition to other states is certainly much smaller and therefore neglected here. Using Landau-Zener formula [\oncite{landau}], we can estimate the transition rate to be $\gamma_2=f_{\rm ave}e^{-2\pi\Gamma}$ with the transition parameter,
\be
\Gamma\equiv\frac{|V_{12}|^2}{\left|dE_{12}(r)/dt\right|_{r_m}}.
\label{tau_2}
\ee
Here $f_{\rm ave}$ is the frequency for two molecules to collide with each other, $E_{12}\equiv E_{|1\rangle}-E_{|2\rangle}$ is the energy difference between the two unperturbed energy eigenstates, and
$V_{12}=\Omega$ is the coupling between them via microwave absorption. Within the semi-classical approximation, we have $dE_{12}(r)/dt=E_{12}'(r)v(r)$, where $v(r)=dr/dt=\sqrt{2(E_b-V_{\rm 2D}(r))/\mu}$ is the local velocity. Here $E_b$ is the bound state energy and $\mu=m/2$ is the reduced mass. By estimating $f_{\rm ave}^{-1}=\int_{r_1}^{r_2}\frac{dr}{v(r)}$ for the two turning points at $r_1$ and $r_2$, we can calculate the lifetime $\tau_2=\gamma_2^{-1}$ directly. In Fig. \ref{loss}(b) we show the full calculation of the transition parameter, $\Gamma$, as a function of $\Omega$ for SrO ($B\sim 10^{10}$Hz and $r_B\sim 11$nm). The typical value of $\Gamma$ is always larger than 5 for SrO, and therefore the obtained life time ($\tau_2$) is very long. We also get $\tau_2\sim 0.1$ sec for KRb molecule ($B\sim 10^9$Hz and $r_B\sim 3.5$nm, if omitting its fine structure).

In order to understand the life time in more details, we can qualitatively estimate its order of magnitude when the bound state is just formed ($E_B\to 0$). First we have $f_{\rm ave}\sim \delta v/2\delta r\sim m(\delta v)^2/2\sim E_K$, where $\delta r$($\delta v\sim 1/m\delta r$) is the uncertainty of radial position(velocity) and $E_K\sim V_m$ is the kinetic energy. Since $E_{12}(r)\to \Delta$ as $r\to\infty$ and changes within the length scale of $r_m\sim \delta r/C$ with a field-dependent constant $C$, we then have $E_{12}'(r)\sim \Delta/r_m$ and therefore $\Gamma\sim C\left|\Omega\right|^2/\Delta E_K$. For example, for the parameter of SrO with $\Omega/B \sim 10^{-4}$ and $\Delta/B \sim 10^{-3}$, we have $C\sim 1$ (see Fig. \ref{loss}(a)) and hence a very long life time as calculated above ($E_K/B \sim 4\times 10^{-7}$ and $\Gamma\sim 23$). We note that, different from the situation in the continuous state, the life time of a super-molecule is longer for stronger dipole moment because the length scale, $r_B=(d^2/B)^{1/3}$, as well as the radial size of the super-molecules, also becomes larger, reducing the relative collision velocity between two molecules. This is consistent with the fact that the adiabatic approximation is more accurate for particles with larger mass. Similar analysis can be also applied to the trimer case. 

The results presented here open a new area for studying few-body physics as well as many-body physics with ultracold molecules. In usual atomic gases, Feshbach molecule is the highest energy bound state with many lower energy states far below. As a result, the whole system of the same species is usually unstable to collapse due to the three-body inelastic collision in the $s$-wave ($p$-wave) channel for identical bosons (fermions). This fact makes the experimental realization of a $Z_2$ quantum phase transition [\oncite{Z2}] between a condensate and a pair condensate of bosons near a Feshbach resonance almost impossible. Similar conclusion also applies to the $p$-wave crossover, or the $p_z$-to-$p_x+ip_y$ transition, predicted in polarized fermion systems [\oncite{fermion_pwave}]. In the polar molecule system we discuss here, however, the ground state super-molecules is long-lived almost without two- and three-body loss near the threshold of dimer formation, suggesting a much more feasible scenario to study important many-body physics such as the phases mentioned above. Although larger clusters may exist when the density is higher, above statement and prediction should still apply near the threshold of dimer-trimer formation. Much more complicated analysis will be needed to fully address this issue.

In summary, we show that the shapes of the 2D effective potentials between polar molecules can be designed to make long-lived super-molecules. This creates a feasible system for studying novel few-body physics as well as many-body problems.

We appreciate fruitful and critical discussion with G. Shlyapnikov, J. Bohn, C. Greene, G. Pupillo, D.V. Fedorov, and A.S. Jensen. This work was supported by NSC (Taiwan).
 

\begin{thebibliography}{99}
\bibitem{Gora_work}
Such kind of bound state has also been mentioned in Ref. [\oncite{shlyapnikov}] in the context of scattering problem, but its general properties are not studied in details there.
\end{thebibliography}
\end{document}